# NUCLEAR MAGNETOHYDRODYNAMIC EMP, SOLAR STORMS, AND SUBSTORMS

**Mario Rabinowitz**
Electric Power Research Institute ; Palo Alto, CA 94303, USA
Inquiries to: Armor Research; 715 Lakemead Way, Redwood City, CA 94062, USA
Mario715@earthlink.net

**A. P. Sakis Meliopoulos and Elias N. Glytsis**
School *of* Electrical Engineering, Georgia Institute *of* Technology, Atlanta, GA 30332, USA

**George J. Cokkinides**
Electrical Engineering Department, University *of* South Carolina, Columbia, SC 29208 , USA

**Abstract**

In addition to a fast electromagnetic pulse (EMP), a high altitude nuclear burst produces a relatively slow magnetohydrodynarnic EMP (MHD EMP), whose effects are like those from solar storm geomagnetically induced currents (SS-GIC). The MHD EMP electric field $E < 10^{-1}$ V/m and lasts $< 10^2$ sec, whereas for solar storms $E > 10^{-2}$ V/m and lasts $>10^3$ sec. Although the solar storm electric field is lower than MHD EMP, the solar storm effects are generally greater due to their much longer duration. Substorms produce much smaller effects than SS-GIC, but occur much more frequently. This paper describes the physics of such geomagnetic disturbances and analyzes their effects.

## 1. Introduction

Electromagnetic pulse (EMP) effects were observed in conjunction with the first nuclear detonation produced by man - the Trinity test of 1945 in New Mexico. At least two kinds of EMP have been identified and studied extensively. Both are related to a high altitude burst, and as such are called HEMP. One is an extremely fast pulse with a duration $\sim 10^{-6}$ sec occurring at the beginning of a burst, which we call TEMP (tachy[fast]EMP). Since the TEMP was thought to have an electric field peak $\sim 50$ kV/m, $\sim 10^6$ times greater than the much slower magnetohydrodynamic EMP (MHD EMP), it was expected that the TEMP would have a much bigger impact; and almost all the attention was focussed on TEMP. TEMP was anticipated to have devastating effects in electrically blacking out huge land masses; and much media coverage was given to this possibility. However, Rabinowitz showed that the impact of TEMP would be local rather than continental, and not be much greater than that of lightning.[1-3] Later independent work by Millard, Meliopoulos, and Cokkinides [4] reached similar conclusions. Consideration of the radiation reaction force [5] and subtle relativistic effects [6] indicate that 50 kV/m may not even be achievable for the preponderance of nuclear weapons that exist.

Therefore for completeness, we shall now focus on the relatively slow MHD EMP which occurs $\sim$ sec after the burst, and has a duration $< 10^2$ sec. We shall see that the effects of MHD EMP are similar to, but generally less severe than those of solar storm geomagnetically induced currents (SS-GIC) which last $> 10^3$ sec, and sometimes continue for days. Both phenomena cause flow of very low frequency current (almost direct) in the earth, and transfer or induce this current to nearby structures. We shall analyze the effects of the voltages and currents that they produce. In addition we will consider substorms which occur on a daily basis, but whose effects at the earth's surface are much smaller than the much less frequent solar storms.

## 2. Solar Burst EMP

### 2.1. Solar storms

The sun emits ionized particles into space on both a steady and a transient basis by means of what is called the solar wind. The steady interaction of the solar wind with the earth's ionosphere and geomagnetic field has no adverse effect on electric power networks. A much stronger and potentially adverse transient constituent results from sunspot activity when the blast wave from a solar flare hits the earth's magnetosphere. This produces electrical system problems about every eleven years during the peak of the sunspot cycle.

Interactions of the earth's magnetic field with the solar wind give rise to auroral currents or aurora electrojets. [7, 8] These high altitude currents produce variations in the earth's magnetic field that-are termed geomagnetic storms. The strength and severity of the geomagnetic storms are strongly related to solar storms, solar flares, coronal holes, and disappearing solar filaments. During geomagnetic storms, the time varying magnetic field can induce electric potential gradients called earthsurface-potentials (ESP) whose magnitude depends on the severity of the geomagnetic storm and on the earth's conductivity. [9, 10]

Due to the 93 million mile journey that the ions, travelling 4 million miles/hr ($2 \times 10^6$ m/sec), must take to reach the earth, the effects of a major solar flare are felt here approximately a day after it erupts. High sunspot activity is usually followed by magnetic storms on the earth. Because of this relatively long time delay, early warning systems have been considered. One system would use an orbiting satellite to give us advance warning to avert perturbative and potentially destructive effects on electric power systems from solar storms which are likely to occur in the sunspot cycle maximum.

The biggest effects in our hemisphere are felt at northern latitudes. In addition, the effects are exacerbated if the ground is poorly conducting in the vicinity of the electric utility network, causing more current to flow through the conductors connected to ground (at more than one point). Good conducting ground helps to reduce the effects on the network, by diverting the induced currents. The magnetic storm may have high activity producing an electric field $\sim 10^{-2}$ V/m for periods ~hour, with intermittent quiescent periods for a duration of about 24 hours as shown in Fig. 1.

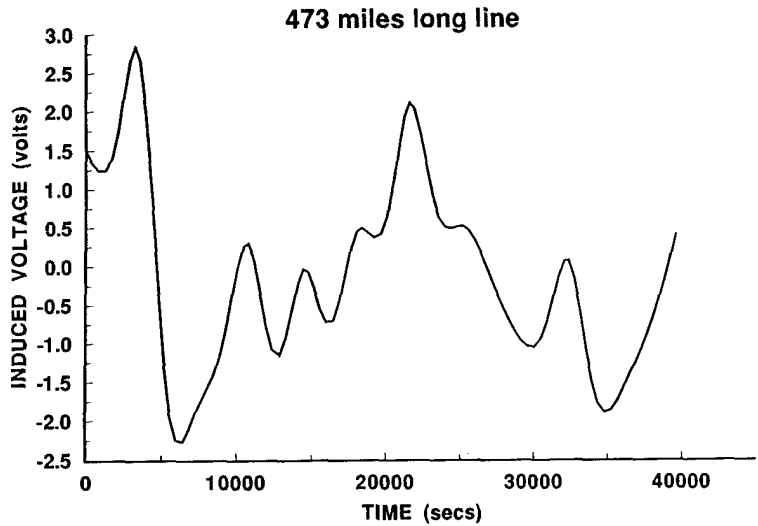

Fig. 1. Induced voltage on a 473 mile (368 km) overhead transmission line from a typical small solar storm - orders of magnitude lower than from a severe storm. Raw data is from a magnetogram recording on May 12-13, 1989 at the Magnetic Observatory Fiirstenfeldbruck.

Slothower and Albertson [11] estimate that "earth potentials of 5 to 10 volts per mile are sufficient to produce direct currents in transformer windings of great enough magnitude to cause saturation, . . . ". The saturation of transformer cores for long enough periods of time can cause them to overheat. This can decrease transformer life. In rare occasions, very large and very expensive transformers have been permanently damaged during periods of very high solar storm activity. It is therefore important to determine if the much shorter duration but higher field MHD EMP could possibly have equally serious effects.

Models that have been developed for the estimation of the SS-GIC differ in the representation of the auroral currents and the earth's conductivity. Sometimes the auroral currents are assumed to be at an infinite distance from the earth's surface. In this case, the perturbed geomagnetic field is modeled as an electromagnetic plane wave [4] as an upper limit calculation. The earth's conductivity is also difficult to model due to the earth's inhomogeneity. [12] The simplest model assumes a flat earth of a uniform effective conductivity, while more sophisticated models include multiple layers of differing conductivities. However, the auroral currents are known to be at high altitudes (100-300 km) and can be modeled like current sheet sources or current line sources (along the east-west direction) of infinite extent above the earth's surface [4,13,14] that is assumed to be flat. The current sheet model gives an upper limit of the induced electric field while the current line model gives a lower limit. The general methodology used for the solution of these problems is based on Price's analysis. [15] More sophisticated studies include Gaussian modeling [16] of the electrojets, or a nonsymmetrical [17] auroral current distribution.

Independently of the model used, the complexity of the physical effect of the auroral electrojets is difficult to represent. In addition, all models assume sinusoidal auroral currents. However, the spectral content of the auroral currents is not known. Due to the above mentioned reasons, the geomagnetic field is usually measured in several positions in the areas of interest using magnetometers. By measuring the horizontal component of the magnetic field, the induced horizontal (perpendicular to the horizontal magnetic field) electric field can be roughly estimated using the plane wave model of Pirjola. [10]

$$E(t) = -\frac{1}{\sqrt{\pi\sigma\mu_0}} \int_0^\infty \frac{g(t-u)}{\sqrt{u}} du$$

$$\approx -\frac{1}{\sqrt{\pi\sigma\mu_0}} \left\{ \frac{4}{3}g(t) + g(t-d) + \frac{2}{3}\sum_{j=1}^{L}(1+a_j)\frac{g[t-(j+1)D]}{\sqrt{j+1}} \right\}\sqrt{D}, \quad (1)$$

where $E(t)$ is the horizontal component of the induced electric field, $\mu_o$ is the permeability of free space, $\sigma$, is the earth's conductivity, g is the time derivative of the horizontal component of the magnetic field, D is the data time-interval, $a_j = 0$ when j is even and 1 when j is odd, and L is the total number of data points that are included in the calculation of $E(t)$. An example, typical of the great majority of small solar storms, of induced voltage from the horizontal component of electric field is shown in Fig. 1, as calculated from Eq. (1). Severe solar storms produce voltages that are orders of magnitude higher.

## 2.2. Substorms

Substorms are a less well known and weaker (with respect to effects on the earth) phenomenon than solar storms and MHD EMP, but have the advantage of much more frequent occurrence. With sensitive enough instrumentation, we might be able to use substorms on a daily basis to study the effects of solar storms and MHDEMP. This would have the advantage of not having to wait for long periods of time before observing an event. Substorms are magnetic storms that occur a few times every day in the center of the earth's magnetosphere. Substorms last about an hour. This is less than the period of one day or more for the relatively infrequent high activity geomagnetic storms.

Substorms are the means by which vast amounts of stored magnetic energy ~$10^{15}$ Joules are released on a daily basis. Substorms may be driven internally by the stored energy in the magnetotail, however most of the time they are driven by the solar wind. This is a comet-like wake ~ $10^2$ earth radii ($R_e$ = 6.4 x $10^6$ m) long, which results from the solar wind's interaction with the earth's magnetic field. A giant magnetic bubble is formed which makes the MHD EMP bubble appear miniscule by comparison. Instead of the MHD EMP bubble radius of hundreds of miles, a gigantic bubble of hot plasma 5 x $10^8$m (300,000 miles) long, $10^6$m (80,000 miles) wide, and 8 x $10^7$ M (50,000 miles) high, as depicted schematically in Fig. 2, is created by the interaction of the solar wind conducting plasma which is threaded and held together by loops of magnetic flux. [18] This plasmoid structure is catapulted as magnetic energy is converted into plasma motion at 5 x 10' m/sec (1,080,000 miles/hr).

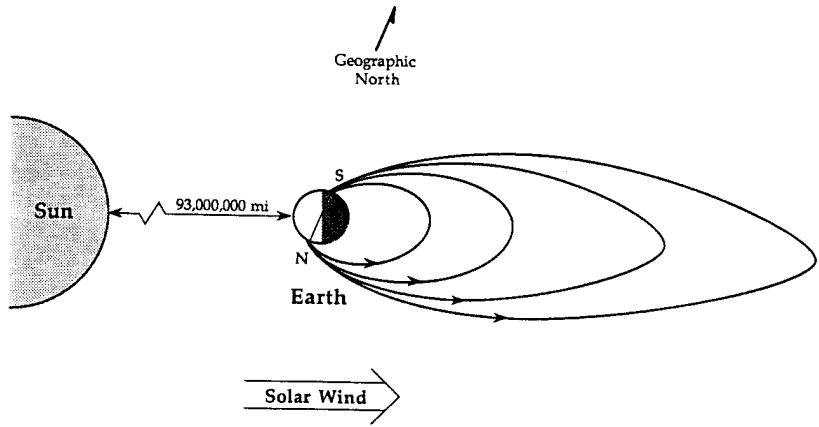

Fig. 2. Gigantic substorm magnetic bubble created by the interaction of the solar wind with the earth's geomagnetic field.

## 3. Nuclear Burst Magnetohydrodynamic EMP (MHD EMP)

There are two kinds of MHD EMP which are approximately similar in their effects, but are caused by physically different perturbations of the geomagnetic field. Though different in their origin, they have much in common with solar storm and substorm disturbance of the earth's geomagnetic field. We shall look a little closer at the physics of this perturbation.

### 3.1. *Description of magnetic bubble EMP (BEMP)*

Of the two kinds of MHD EMP, we call the first one BEMP (the B stands for "bubble"), in which a magnetohydrodynamic bubble is produced as the nuclear bomb's ionized debris expands. The high temperatures and copious x-ray emission of a nuclear burst produce vaporization and ionization of the bomb material. This moves out as a quickly expanding plasma shell with an initial velocity of $\sim 10^6$ m/sec. If the shell were not conducting, it would simply intercept and enclose more and more of the earth's magnetic flux as it expands. However, because it is conducting, currents are set up in this shell whose magnetic flux tends to cancel the earth's flux in accordance with Lenz's law. Even though the shell gets much bigger, its own currents try to limit the flux to the small amount initially inside the small shell. This is the normal diamagnetic response of a conductor that is exposed to an external magnetic field.

The magnetic field will start to diffuse in as governed by Maxwell's equations whose time dependent solution [19] is

$$\frac{\partial B}{\partial t} = \frac{1}{\sigma\mu}\nabla^2 B - \frac{\varepsilon \partial^2 B}{\sigma \partial t^2} - \frac{1}{\sigma\mu}\nabla[\nabla \cdot B] , \qquad (2)$$

where $B$ is the magnetic flux density, $\sigma$ is the conductivity, $\mu$ is the permeability, and $\varepsilon$ is the permittivity. The last term in Eq. (2) is 0, since there are no magnetic monopoles in this problem. The second term is 0 in the absence of displacement currents. Thus Eq. (2) reduces to the standard diffusion equation

$$\frac{\partial B}{\partial t} = \frac{1}{\sigma\mu}\nabla^2 B . \qquad (3)$$

Solution of Eq. (3) gives the diffusion time constant for a magnetic field into a medium of permeability μ and conductivity σ:

$$\tau = \frac{\mu \sigma \delta^2}{2},  \quad (4)$$

where δ is the penetration depth. Separate knowledge of μ, σ, and δ are not needed to determine τ. Since

$$\delta = [2/\mu\sigma\omega]^{1/2}$$

(ω is the angular frequency), this let's us find τ quite easily: τ = 1/ω. The geomagnetic field will be kept out as long as τ is long compared to the time for the external magnetic pressure to build up. We will make an approximate calculation of this.

As this magnetic bubble grows, it excludes or pushes away the earth's magnetic field, yielding a concentration of flux outside of it and almost no flux inside it, as shown in Fig. 3. As the magnetic flux density, *B*, increases outside the bubble, the magnetic pressure $B^2/2\mu$, and damping by the air viscosity act to slow down and finally stop the expanding magnetic fireball. This occurs when the magnetic pressure equals the kinetic pressure, and the magnetic field goes back inside the bubble, producing an Alfen wave.

An initially spheroidal fireball of ionized matter probably distorts into a prolate spheroid as it expands due to the disproportionate enhancement of the magnetic field in the equatorial region of the bubble. In the far field, the effects that propagate to the earth should decrease in strength rapidly as the distance to the source region increases. The maximum field is quite small, only ~ $10^{-1}$ V/m with a period from 2 to 100 sec. This pulse occurs about 2 to 5 seconds after the nuclear explosion. Even small anti-satellite bursts at very high altitudes may produce MHD EMP effects.

*3.2. Analysis of magnetic bubble EMP (BEMP)*

Some simple calculations will shed light on the size of the magnetic bubble, the radiated power output, the radiation efficiency, the maximum power density, and the magnitude of the radiated electric field from BEMP. This can enable us to check the self-consistency of the various parameters, and determine upper limits on them if we wish.

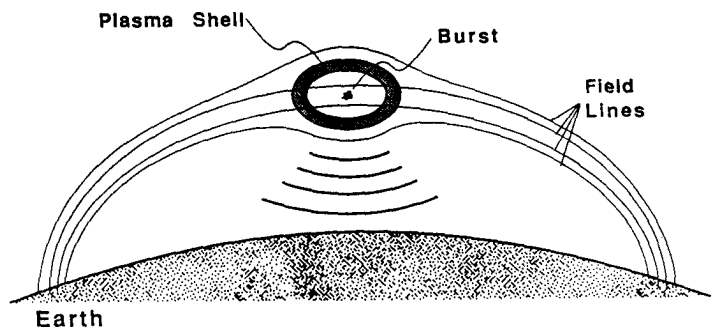

Fig. 3. Magnetic bubble in the geornagnetic field produced by a nuclear burst causing the BEMP form of MHD EMP.

In producing BEMP, the expanding ionized (conducting) shell pushes the geomagnetic field out of its way. The bubble formed in the geomagnetic field reaches its maximum size when the energy in the <u>excluded field</u> equals the initial kinetic energy, T, of the conducting shell neglecting viscosity effects such as air damping. This approach allows use of the initial unperturbed value of the geomagnetic flux density, B, rather than its larger compressed value which could be derived by equating magnetic and kinetic pressures when the bubble stops expanding. This latter approach could be done iteratively to get a self-consistent solution.

In this first rough calculation, assume that the bubble does not break up into smaller bubbles, that it maintains a spherical shape without distortion, and that atmospheric damping may be neglected.

$$\int_0^v \frac{B^2}{2\mu} dV = \frac{\langle B^2 \rangle}{2\mu} \int_0^v dV$$

$$= \frac{\langle B^2 \rangle}{2\mu} \left( \frac{4}{3} \pi R^3 \right) = T \quad (5) \, \& \, (6)$$

{$B^2$}: mean squared value of flux density of the unperturbed geomagnetic field inside the volume of the bubble. Take {B}~ 0.5G = 5 x 10-5 Wb/m². V: maximum volume of the bubble when it stops expanding. μ: permeability of medium. R: radius of the bubble. T: initial kinetic energy of the conducting debris of the bomb. Because ~ 70% of the bomb's energy goes into x-rays, one may expect T < 30% of the bomb's yield.

Solving Eq. *(5)* for R yields:

$$R = \left[ \frac{3\mu T}{2\pi \langle B^2 \rangle} \right]^{1/3} . \quad (7)$$

For a 1.4-Mt bomb, $T = 0.3(1.4 \, \text{Mt})(4.2 \times 10^{15} \, \text{J/Mt}) = 1.76 \times 10^{15}$ J, which implies

$$R = 7.5 \times 10^5 \, \text{m} = 468 \, \text{miles} . \quad (8)$$

This bubble radius is greater than the height of most bursts. Actually, as we shall see, the bubble becomes a prolate spheroid and can thus miss the ground.

Taking 100 sec as the given pulse duration, the average power radiated is

$$\langle P \rangle = fT/t$$
$$= f(1.76 \times 10^{15} \, \text{J})/100 \, \text{sec}$$
$$= (1.8 \times 10^{13} \, \text{W}) f \quad (9)$$

where $f < 1$ is the average conversion efficiency of the kinetic energy into radiation.
The power density near the bubble is:

$$\langle P \rangle / 4\pi R^2 < (1.8 \times 10^{13} \, \text{W}) f / 4\pi (7.5 \times 10^5)^2$$
$$= (2.5 \, \text{W/m}^2) f . \quad (10)$$

The maximum radiated power density at points far from the bubble will be much lower than this value as the power density decreases with increasing distance.

To estimate the BEMP-generated electric field near the bubble from the Poynting vector and Eq. (10):

$$|\mathbf{E} \times \mathbf{H}| = E^2/z, \; z = 377 \Omega$$
$$= \langle P \rangle / 4\pi R^2 = (2.5 \text{ W/m}^2) f \; . \tag{11}$$

Equation (11) yields the result

$$E = 30 f^{1/2} \text{ V/m} \; . \tag{12}$$

Two approaches may be taken in estimating the average conversion efficiency $f$. One is to use the maximum electric field, E, which is given to be 0.1 V/m. This implies $f \sim 10^{-5}$. The other one is to note that a period of 100 sec corresponds to a wavelength $\lambda$ of $3 \times 10^{10}$ m. The length of the radiating antenna is $\sim R = 7.5 \times 10^5$ m from Eq. (8). When the antenna is so short compared with the wavelength, the efficiency is roughly

$$f \sim R/\lambda = 2.5 \times 10^{-5} \sim 10^{-5} \; . \tag{13}$$

We may now use this value of f to determine the average radiated power, $\langle P \rangle$, and an upper limit on the power density near the bubble, $\langle P \rangle / 4\pi R^2$ from Eqs. (9) and (10).

$$\langle P \rangle \sim 10^8 \text{ W} \; . \tag{14}$$
$$\langle P \rangle / 4\pi R^2 \sim 2.5 \times 10^{-5} \text{ W/m}^2 \; . \tag{15}$$

Due to a greater compression of the magnetic field lines in the equatorial region of the bubble, its shape will become somewhat like a prolate spheroid (ellipsoid), as shown in Fig. 3. For a diamagnetic sphere, the field is increased most at the equator, and is 1.5 times greater at the equator than the unperturbed external field. For a diamagnetic cylinder in a transverse field, the field is 2 times larger at the equatorial lines than the unperturbed external field. As the expanding sphere distorts into a prolate ellipsoid, the field enhancement is (1-D)` where for an eccentricity e, the demagnetizing factor is

$$D = (e^{-2} - 1)[(2e)^{-1} \ln\{(1+e)/(1-e)\} - 1] \; . \tag{16}$$

By an iterative process, one can calculate the shape of the resulting ellipsoid. For simplicity, let's assume that the ratio of the major semiaxis a, to minor semiaxis b, is about 2 to 1. Equation (6) yields the following:

$$[\langle B^2 \rangle / (2\mu_0)] V = [\langle B^2 \rangle / (2\mu_0)][4\pi a b^2/3] = T \; . \tag{17}$$

Solving Eq. (17) for $b$,

$$b = [3\mu_0 T / 8\pi B^2]^{1/3} = 6 \times 10^5 \text{ m} = 372 \text{ miles} \; , \tag{18}$$

and

$$a = 2b = 1.2 \times 10^6 \text{ m} = 744 \text{ miles} \; . \tag{19}$$

Thus, the magnetic distortion of the bubble may keep the bottom of it from hitting the ground for a high enough altitude shot. Of course, viscosity effects such as air damping will act to reduce the size of the bubble and to shift the direction of the expansion somewhat more away from the earth where the air density decreases.

### 3.3. Atmospheric heave EMP (AEMP)

A second kind of magnetic perturbation is also called MHD EMP, and is similar to the first. It will be called AEMP here to distinguish it from BEMP. AEMP is caused by the heave of bomb-heated ionized air across the geomagnetic field, and occurs more than 10 seconds after a bomb burst. Figure 4 shows the formation of a small magnetic bubble and the beginning of the atmospheric heave for a low-altitude burst. About 70% of the energy released by a bomb appears as x-rays that photo-ionize the air. This process forms large ionospheric current loops with mirror images in the Earth as illustrated in Fig. 5. The effects of perturbing the geomagnetic field extend out more than $10^6$m from the source point and last for $\sim 10^2$ sec. Both the field and frequency are very low, at 0.001 to 0.03V/m and 0.01 Hz.

### 3.4. Discussion

In addition to instrumenting utility networks to study the effects of solar storms every eleven years, it may be more feasible to instrument a test line to study the effects of substorms on a daily basis. With proper scaling, it may be possible to extrapolate the effects to SS-GIC and MIID EMP. Fortunately, substorms have negligible effects on the earth. Otherwise their daily occurrence would be quite troublesome. If we can determine that the effects of MHD EMP are no more severe than SS-GIC, then we have a relatively well known base for comparison of the probable effects. It is not presently known whether the long duration SS GIC or the higher amplitude, but significantly shorter duration MHD EMP presents the more severe stress for an electric utility network.

It is likely that the SS GIC produces transformer core saturation for much more than $10^{-1}$ of the ac cycle since transformer cores are operated at from 50 to 90% of saturation (at the maximum magnetization current), to minimize overall costs. Thus it appears that a higher electric field MHD EMP acting for only $10^{-2}$ of the time duration of SSGIC will probably not be as severe - provided that the induced slowly varying current is small compared with the normal line current so that the major effects are due to the offset (biased) line current.

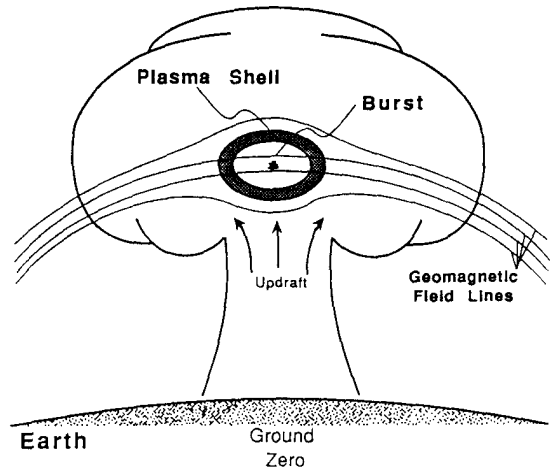

Fig. 4. Formation of a small magnetic bubble from a nuclear burst at the beginning of the AEMP (atmospheric heave EMP) form of MHD EMP.

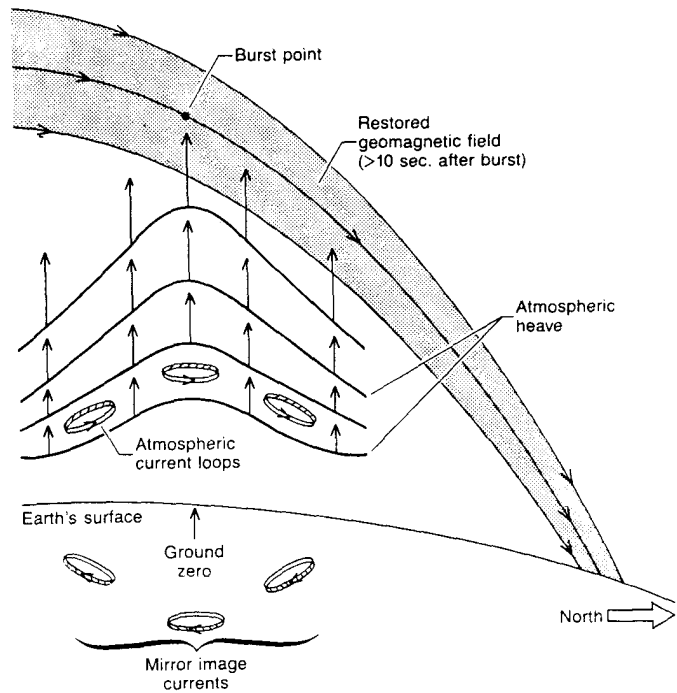

Fig. 5. AEMP ionospheric current loops with magnetic moment mirror images in the earth.

The specific details of latitude, electric field magnitude and duration, ground conductivity, line orientation, transformer design, etc., make the problem sufficiently complicated that a thorough analysis is necessary, which follows.

## 4. Examination of the Relative Effects of MHD EMP and Solar Storms

Although Rabinowitz [20] made a brief analytical comparison, a comprehensive comparison of the effects of MHD-EMP and solar storms (SS) has not previously been done. Our work addresses the effects of these geomagnetic disturbances on a comparative basis. For this purpose, a system approach has been adopted to analyze these problems. We simulated a power transmission system consisting of transmission lines, transformers, and all associated grounding structures exposed to MHD EMP or SS excitation. A model of an electric power transmission line which takes into account MHD EMP/SS geomagnetic coupling was developed. The form of the transmission line model is in terms of a multiple input-multiple output linear system. Since the coupling of MHD EMP or SS induced voltages to power lines is mainly through the line grounding system, it is important to accurately model the power line tower grounding as well as the terminal substation grounding system. For this purpose, the EPRI grounding models developed by Georgia Tech were used. Power transformers were modeled with their nonlinear magnetization characteristics. Using this model, system studies have been performed to determine transformer magnetization currents and reactive power requirements. The example system utilized in the study is a simplification of an actual system provided by Minnesota Power Company.

The effects of the MHD EMP on power system grids are very similar to those of geomagnetic storms. Exact models for the MHD EMP calculations do not exist to our knowledge. Estimates of the induced electric field can be calculated as indicated in Sec. 3.3, or obtained from nuclear test data. The most valuable information about MHD EMP is empirically known from the magnetometer data acquired during nuclear tests. The design waveform for the simulation of MHD EMP is the one measured during the Starfish nuclear test. The induced electric field shown in Fig. 6 was derived by sampling and linear interpolation from the magnetometer data of Legro et al.[21] Rackliffe et al. [22] also looked at the effects of MHD EMP, as did Klein et al. [23]

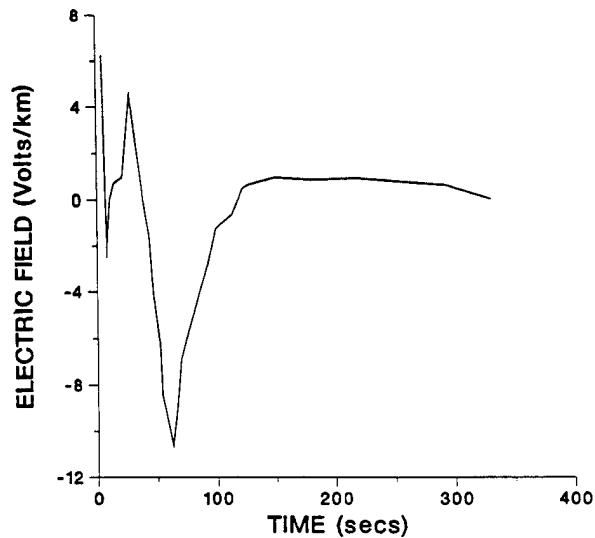

Fig. 6. Horizontal component of the induced electric field from MHD-EMP as derived from data in Ref. 22

We will describe the power delivery system model for the study of geomagnetic disturbances. The model is based on a time domain simulation algorithm similar to the EMTP (ElectroMagnetic Transients Program). Each power system element is modeled with a set of differential equations which are solved in the time domain. For the study of geomagnetic disturbances, two power system elements are very important: (I)iron core transformers and (2) long transmission lines. Specifically, the grounding of the transformer neutrals and interconnecting long transmission lines provide the gate for geomagnetically induced currents to enter the power system. Magnetic core transformers reach saturation when geomagnetically induced currents flow in their windings and cause most of the undesirable effects. This section describes in detail these two models.

**4.1. Saturation of transformer cores**

The nonlinear magnetization characteristics of iron core transformers are represented in the transformer model utilized in this study. The modeling procedure is illustrated with a simple case of a single phase transformer shown in Fig. 7a. Most of the magnetic flux circulates through the core and thus links both windings. A part of the flux, however, leaks from the magnetic core and links only one winding. These flux paths are represented by the equivalent magnetic circuit illustrated in Fig. 7b. Specifically, the reluctances, $R_1$, $R_2$, and $R_3$ represent the path of the leakage flux, while the reluctances $R_4$ and $R_5$ represent the path of the flux through the core (magnetizing flux). Reluctances $R_4$ and $R_5$ are nonlinear, since their value depends on the flux magnitude. From the magnetic circuit of Fig. 7b, the electric equivalent circuit of Fig. 7c is derived. The leakage reluctances, $R_1$, $R_2$, and $R_3$, are represented by the inductances $L_{1u}$ and $L_{2u}$ and the magnetization reluctances $R_4$ and $R_5$ are represented by the magnetizing inductance $L_3$. The leakage inductances are linear devices, and thus they are represented by inductors of a specific constant inductance. However, the magnetizing inductance is nonlinear, and its representation is based on expressing its flux linkage as a nonlinear function of the electric current. The equations that describe the transformer equivalent circuit are as follows:

$$V_{1u} = R_{1u}i_{1u} + L_{1u}di_{1u}/dt + d\lambda_u/dt$$

$$V_{2u} = R_{2u}i_{2u} + L_{2u}di_{2u}/dt + d\lambda_u/dt \tag{20}$$

where: $\lambda_u = g(i_{1u} + i_{2u})$, and $g(i)$ is the nonlinear magnetization curve.

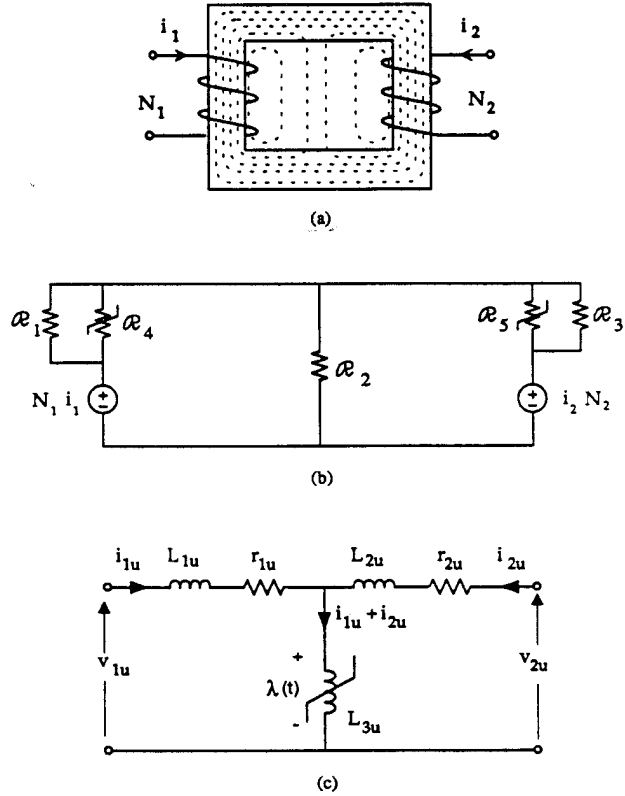

Fig. 7. (a) Schematic representation of a single phase transformer. (b) Equivalent magnetic circuit. (c) Equivalent electric circuit.

The leakage inductance is assumed to be constant independent of the saturation level of the transformer. This is not exactly true but the influence of the leakage inductance on the level of geomagnetically induced currents is secondary and negligible. This has been observed with the overall model where even doubling of the leakage inductance did not alter the rise time or level of the geomagnetically-induced currents. As a result, the iron core transformer model of Fig. 7c is appropriate for the purposes of this study. Three phase transformer banks are represented with three single phase transformers, appropriately connected (wye-delta, etc.). Three phase transformer models are derived with a procedure similar to the one described for single phase transformers by modeling the magnetic circuit and subsequently deriving the input/ouput model in terms of electrical quantities, i.e. voltage and current. We did not use three phase transformer models in this study.

### 4.2. Transmission line system

A time domain state space model based on the methodology developed by Cokkinides and Meliopoulos is used to model the transmission line system. [24] It is capable of representing effects of geomagnetically induced currents (GIC), as well as transmission line parameter frequency dependence, and line tower grounding. The transmission line model involves two components: overhead conductors and earth return; and grounding system. Consider the typical transmission line shown in Fig. 8. The section of overhead conductors between any two consecutive towers comprises a "conductor component" while each tower with its grounding system is a component of the grounding system. Each component is modeled by an equivalent admittance matrix (which is a function of frequency) and equivalent current sources. Then, using nodal analysis, the equivalent circuit of the entire transmission line and GIC coupling is formed. The resulting model is in the form of a passive circuit of known admittance matrix and lumped current sources connected at the line terminals. This model is finally converted to the time domain using Fourier techniques. The derivation of the conductor and grounding system equivalent circuits are presented in the following sections.

### 4.3. Elevated conductors

An overhead transmission line conductor section, in the presence of geomagnetically induced currents, is represented by the equations:

$$\frac{\partial v(x,t)}{\partial x} = -Ri(x,t) - L\frac{\partial i(x,t)}{\partial x} + Uv_g(x,t)$$

$$\frac{\partial i(x,t)}{\partial x} = -Gv(x,t) - C\frac{\partial v(x,t)}{\partial x}$$

(21)

v: line voltage column matrix ($V_a$, $V_b$, $V_c$, $v_n$) with respect to remote earth (volts). i: line current column matrix $i_a$, $i_b$, $i_c$, $i_n$, (amperes). R: line series resistance matrix (ohms/m). L: line series inductance matrix (H/m). G: line shunt conductance matrix (S/m). C: line shunt capacitance matrix (F/m). $v_g$: component of geomagnetically induced voltage (GIV) in the direction of the line (V/m). U: column matrix, every entry of which is unity.

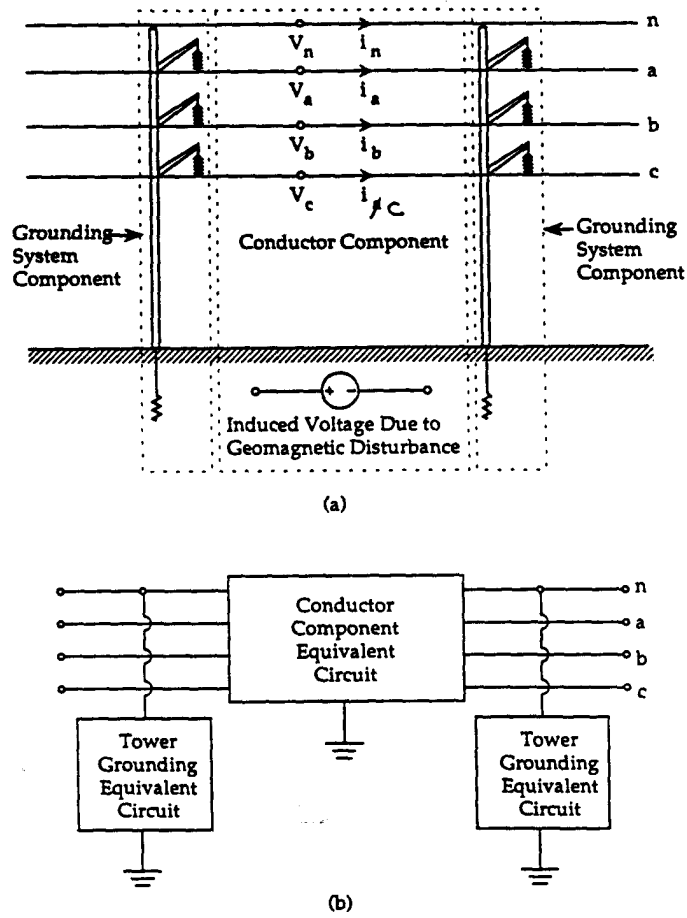

Fig. 8. Typical 3-phase, 4-wire transmission line section: (a) Physical configuration. (b) Equivalent circuit.

The Fourier transforms of Eqs. (21) are:

$$\frac{\partial V(x,\omega)}{\partial x} = -Z(\omega)I(x,\omega) + UV_g(x,\omega)$$
$$\frac{\partial I(x,\omega)}{\partial x} = -Y(\omega)V(x,\omega)$$
(22)

where $Z(\omega) = R(\omega) + j\omega L(\omega)$, and $Y(\omega) = G(\omega) + j\omega C(\omega)$. For the present study, the following three assumptions have been made: 1) for a short line span, Vg is assumed constant with respect to position; 2) the resistance and inductance are computed using Snelson's [25] complex depth of return method; 3) the capacitance matrix is independent of frequency.

A Norton equivalent circuit is constructed using Eqs. (22) as follows: 1) with the use of eigenvalue analysis, Eqs. (22) are transformed into a set of decoupled equations; 2) a general analytical solution is computed for the decoupled equations; 3) a specific solution is obtained by using the voltages and currents at the line ends as boundary conditions; and 4) the solution of the decoupled equation is converted back to the original variables, yielding:

$$\begin{bmatrix} I_1(\omega) \\ I_2(\omega) \end{bmatrix} = \begin{bmatrix} Z^{-1}(\omega) & 0 \\ 0 & Z^{-1}(\omega) \end{bmatrix} \begin{bmatrix} V_g(\omega) \\ V_g(\omega) \end{bmatrix} + \begin{bmatrix} Y_1(\omega) & Y_2(\omega) \\ Y_2(\omega) & Y_1(\omega) \end{bmatrix} \begin{bmatrix} V_1(\omega) \\ V_2(\omega) \end{bmatrix} \qquad (23)$$

$I_1$, $I_2$: matrices containing the current phasors at the ends of the line. $V_1$, $V_2$: matrices containing the voltage phasors at the ends of the line. $V_g$: matrix each entry of which is the geomagnetically induced voltage per unit length. $Z: R(\omega) + j\omega L(\omega)$, line series impedance matrix. $Y_1$, $Y_2$: frequency-dependent matrices derived from containing the voltage phasors at the ends of the line. Equation (23) allows the representation of each transmission line segment by a Norton equivalent circuit. The first term on the right-hand side represents an external current source, while the second term is a shunt admittance term multiplied by the voltage matrix. Note that this is a frequency dependent model, and thus its parameters, must be computed at every frequency of interest.

*4.4. Supporting towers*

Each tower and its grounding structure are represented by a step response. It is defined as the current flowing into the tower from the ground wire support point when a unit step voltage is applied at the same point. The step response of the tower and its grounding system can be determined experimentally as done by Meliopoulos and Moharam [26] or analytically. [25,27] When computed analytically, finite element analysis is utilized to solve for the flow of currents in the earth. Then a convolution algorithm derived by Papalexopoulos and Meliopoulos is utilized to evaluate the tower and ground step response.[27] The tower model has been validated with data obtained by Bonneville Power Administration (BPA).[26] The admittance of the tower and grounding system at any given frequency is computed from the step response with an appropriate Fourier transform.

*4.5. System integration*

The equivalent circuit of the entire transmission line is constructed by combining the equivalent circuits of each conductor section and tower grounding systems. The procedure is based on nodal analysis method, where all internal node voltages and currents are eliminated, and all internal current sources are represented by equivalent circuit sources at the terminals of the line. In order to utilize the developed model in a time domain simulation, the equivalent circuit parameters are transformed into the time domain. Specifically, the admittance matrix of the passive part of the equivalent circuit is transformed to an impulse response matrix, and the equivalent current sources (which are also computed as functions of frequency) are transformed into time domain waveforms. A discrete Fourier transform method is used for this purpose.

The Snelson transformation is applied before Fourier transformation to minimize resulting time domain waveform durations. [27] Specifically, the voltage and current variables are replaced by B and F variables as follows:

$$\begin{aligned} B(\omega) &= \begin{bmatrix} V_1(\omega) \\ V_2(\omega) \end{bmatrix} - G \begin{bmatrix} I_1(\omega) \\ I_2(\omega) \end{bmatrix} \\ F(\omega) &= \begin{bmatrix} V_1(\omega) \\ V_2(\omega) \end{bmatrix} + G \begin{bmatrix} I_1(\omega) \\ I_2(\omega) \end{bmatrix} \end{aligned} \qquad (24)$$

where G is a real appropriately selected $2_n \times 2_n$ matrix (n = number of conductors). Applying the above transformation to Eq. (23) yields:

$$B(\omega) = M(\omega)F(\omega) + A_g(\omega) \qquad (25)$$

where
$$M(\omega) = (G(\omega) + Y(\omega))^{-1}(G(\omega) - Y(\omega))$$

$$A_g(\omega) = -2[Y(\omega) + G(\omega)]^{-1} \begin{bmatrix} Z^{-1}(\omega)V_g(\omega) \\ Z^{-1}(\omega)V_g(\omega) \end{bmatrix}.$$

The matrix M(ω) and the column matrix $A_g$(ω) are next transformed into time domain functions using the FFT algorithm. The function matrices m(t) and $a_g$(t) comprise a time domain model of the entire transmission line with GIV coupling. Specifically, m(t) contains the impulse response of the transmission line (based on Snelson's transformation) and the functions $a_g$(t) represent the GIV effects. These functions are utilized in a convolution based algorithm, in order to simulate the operation of transmission lines with GIV coupling in the integrated power system. This algorithm is described in the following section.

*4.5. Solution methodology*

By means of a discrete convolution procedure, the transmission line model is cast into a resistive companion form. This algorithm allows the model to be interfaced with models of other power system components, thus forming a model of an integrated power system. (This is the standard technique followed by several time domain simulation programs such as the EMTP and the PSTS, Power Systems Transients programs.) Specifically, combining Eqs. (24) and (25) and upon transformation into the time domain, yields $Yv(t) = b(t) + b_g(t) + i(t)$ where

$$Y = G^{-1}, \ b(t) = \mathcal{F}^{-1}[YM(\omega)F(\omega)], \ b_g(t) = \mathcal{F}^{-1}[YA_g(\omega)], \ v(t) = \mathcal{F}^{-1}\begin{bmatrix} V_1(\omega) \\ V_2(\omega) \end{bmatrix},$$

$$i(t) = \mathcal{F}^{-1}\begin{bmatrix} I_1(\omega) \\ I_2(\omega) \end{bmatrix}.$$

The above equation can be solved by discrete time techniques in terms of the impulse response model defined in the previous section. Specifically, let $v_n$ and $i_n$ represent the values of the voltage and current matrices at the line ends at the nth time step. Then:

$$Y v_n = b_{n-1} + b_{gn.} + i_n \qquad (26)$$

where $Y = G[I + S_o]^{-1}[I - S_o]$,

$$b_{n-1} = G[I + S_o]^{-1} \sum_{k=1}^{N}(S_k - S_{k-1})(v_{n-k} - G^{-1}i_{n-k})[I - S_o] \text{ and } b_{gn} = G[I + S_o]^{-1}a_{gn}$$

where $S_i$ represents the transmission line step response (at the *i*th time interval), i.e. it is the integral of the impulse response m(t), performed in discrete time. The above equation is a resistive companion form representation of a transmission line with GIV coupling. Specifically, the real matrix Y is the admittance matrix of a resistive network (Y in Fig. 9). The matrix $b_{n-1}$ represents the past history dependent current sources. The entries of the current source matrix *bn-1* are computed by discrete convolution as shown in Eq. (26). The matrices $b_{gn}$ are the independent current sources ($b_g$, in Fig. 9), representing the effects of GIV.

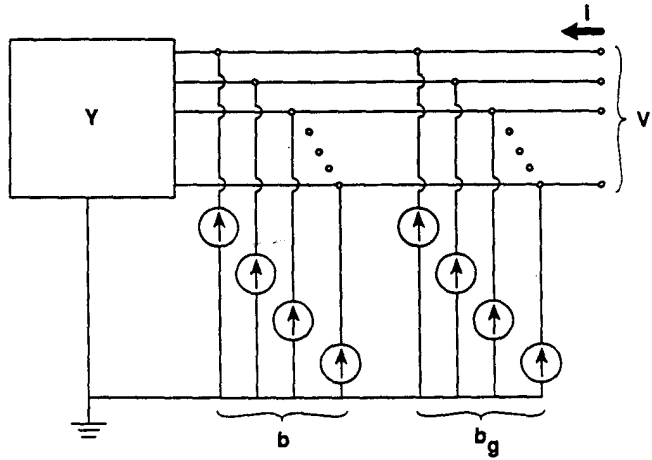

Fig. 9. Transmission line model with coupling to geomagnetically induced currents in the resistive companion form.

### 4.6. The prototype network

Our prototype network is a simplified and modified version of the Minnesota Power Company 500 kV line between Dorsey and Minneapolis.- Specifically it consists of a 500 kV transmission line, terminated by three phase transformer banks at both ends, It is assumed that no intermediate substations exist. Figure 10(a) illustrates a single line diagram of the test system. The transmission line data are listed in Table 1. During normal operation, the ground wires are not electrically connected to the transmission line towers. Tower configuration data specifies the location of the center of each phase bundle and each ground wire with respect to a Cartesian coordinate system with its origin located at the center of the tower base.

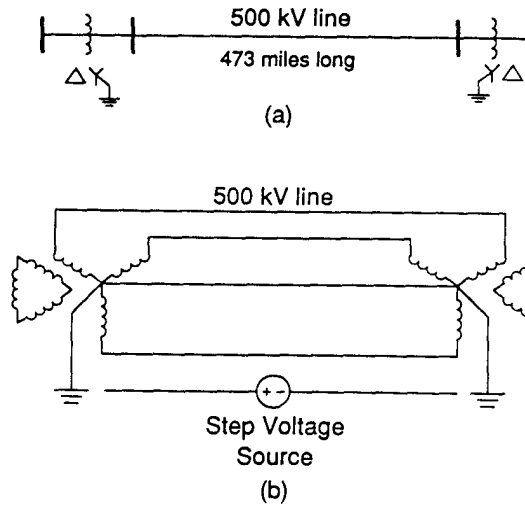

Fig. 10. Test system: (a) Single line circuit. (b) Three phase circuit.

Table 1. Transmission line data.

GENERAL DATA

| Line Length | 473 miles |
| --- | --- |
| Configuration | 3 conductor bundle per phase |
| Bundle Spacing | 18 inches |
| Tower Spacing | 0.25 miles |
| Tower Footing Resistance | 30 ohms |

CONDUCTOR DATA

| Conductor | Type | O.D inches | Resistance (ohms/mile) |
| --- | --- | --- | --- |
| Ground Conductors | 7/16 steel | 0.4375 | 4.435 |
| Phase Conductors | 1192 ACSR | 1.3020 | 0.080 |

TOWER CONFIGURATION DATA

| Conductor | x-coordinate (feet) | y-coordinate (feet) |
| --- | --- | --- |
| Phase A | −32.0 | 97.5 |
| Phase B | 0.0 | 97.5 |
| Phase C | 32.0 | 97.5 |
| Sky Wire 1 | −35.0 | 129.5 |
| Sky Wire 2 | 35.0 | 129.5 |

It is assumed that each of the three phase transformer banks consists of three single phase transformers connected DELTA/GROUNDED Y. As shown in Fig. 10 (b), the grounded Y side is connected on the 500 kV transmission line. The characteristics of each single phase transformer are as follows: voltage = 115/288 kV, power rating = 350 MVA, leakage reactance 0.10 pu, magnetizing current = 0.01 pu, winding resistance (high voltage side) 1.5ohms. The transformer core magnetization characteristics are described by the piecewise-linear function $g$, Eqs. (20).

The effects of GIV to the test system are assessed in two steps. First, the steady state direct current through the transformer is computed for a set of GIV levels and various values of line parameters. Next, the transient response of the transformer excitation current is computed for a set of GIV levels to determine the time constants involved to reach saturation. Finally, the design GIV waveform from MHD EMP and solar storms are applied to the test system to determine the relative effects.

## 5. System Equilibrium Response

We next calculate the equilibrium steady state response of the test system under GIV excitation. In order to gain insight into the system parameters that determine the system behavior under GIV excitation, a simple approach is first used. Specifically, an equivalent dc model of the system is constructed. Using this simple model, the magnitude and distribution of geomagnetically induced currents are evaluated. The dc model of the test system is constructed by considering only the resistances of each system component. Specifically, an equivalent circuit is constructed containing the dc models of the transformers, transmission line, grounding system, and geornagnetically induced voltage. This equivalent circuit is constructed containing the dc models of the transformers, transmission line, grounding system, and geomagnetically induced voltage. This circuit is illustrated in Fig. 11. The equivalent circuits of each component are described next.

The transformers are represented by their winding resistances. Specifically, the windings of the Y connected (high voltage) side of the transformer exhibit three parallel paths to the flow of the electric current injected at the transformer neutral. Assuming that the windings are identical, the equivalent resistance is 1/3 of the winding resistance of each high voltage winding. The transmission line is represented by a dc equivalent circuit. Each of the phase wires is represented by a resistance equal to the total dc resistance of the phase conductor. The neutral wire is represented by its dc resistance. The neutral wire may or may not be multiply grounded. Figure 11 illustrates the tower footing grounding which is represented by its dc resistance. The substation grounds at each line end are represented by 1 ohm resistances, connected from the transformer Y side neutral to remote earth.

Finally, the equivalent circuit of the earth containing the geomagnetically induced voltage is represented by a series of Thevenin equivalent circuits connected between consecutive tower grounds and substation grounds. Thus, for each line segment, a separate Thevenin equivalent of the earth is used. Each Thevenin equivalent consists of a voltage source representing the geomagnetically induced voltage, and a resistor representing the earth path resistance. The earth path resistance is highly dependent on frequency. The earth path resistance computed at 0.6 Hz was used. This model was employed to study the effects of multiply-grounded ground wires on the steady state direct current through the transformer winding. The values used for the parametric study are listed in Table 2.

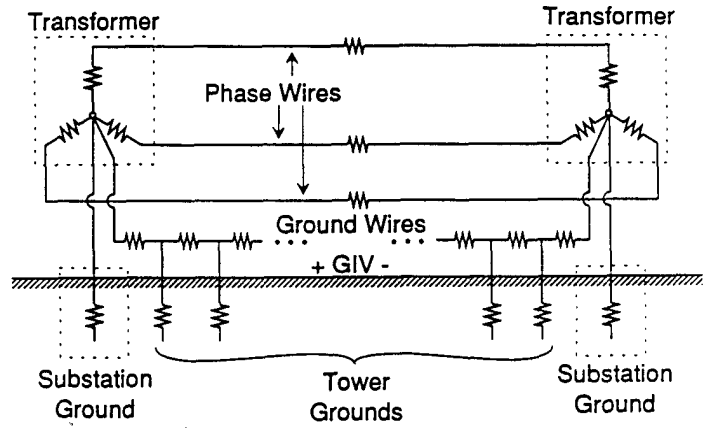

Fig. 11. Direct current equivalent circuit of the test system illustrating the tower footing grounding.

| Table 2. List of parameter values used in the study. | |
|---|---|
| GIV | 1 volt/mile |
| Earth path resistance | 0.001 ohm/raile |
| Tower footing resistance | 5, 30, 100 and infinite ohnis |
| Tower spacing | 0.25 mile |
| Equivalent phase conductor resistance | 0.00889 ohms/mile |
| Ground conductor resistance | 4.435 and 1.240 ohms/niile |
| Total line length | 473 miles |

Figure 12 shows the results of the parametric study which gives the dc current through the transformer winding as a function of the tower footing resistance and for two different ground wire sizes. Note that there is a substantial effect of the line grounding parameters on the steady state value of the direct current through the transformer. For example, for a line with 5 ohm tower footing resistance, and ground wire of 1.24 ohms/mile, the steady state dc current throughout the transformer will be about one-half of what it would have been if the transmission line tower was insulated from the ground wire (37 amperes versus 65 amperes).

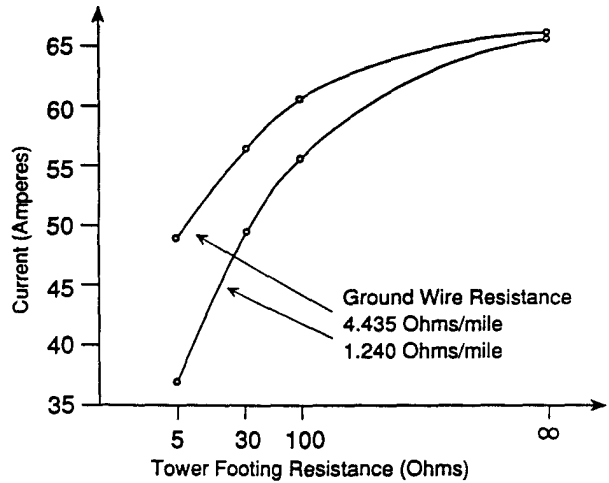

Fig. 12. Transformer steady state direct current for two ground wire sizes.

## 6. Saturation Time Constants

The test system of Fig. 10 was simulated in the time domain to determine the time constants involved to reach steady state operation. Specifically, the time constants were defined as the time required to reach 63% of its steady state value. Figure 13 illustrates a typical simulation and demonstrates the time constant. The figure also illustrates the parameters of the simulation as well. The time constants were computed for the parameter values listed in Table 2. The results of the parametric study are illustrated in Table 3. Note the wide variation of time constants (145 to 2.3 seconds). System parameters drastically affect time constants.

## TYPICAL RESULTS OF TIME DOMAIN SIMULATION
## OF THE TEST SYSTEM

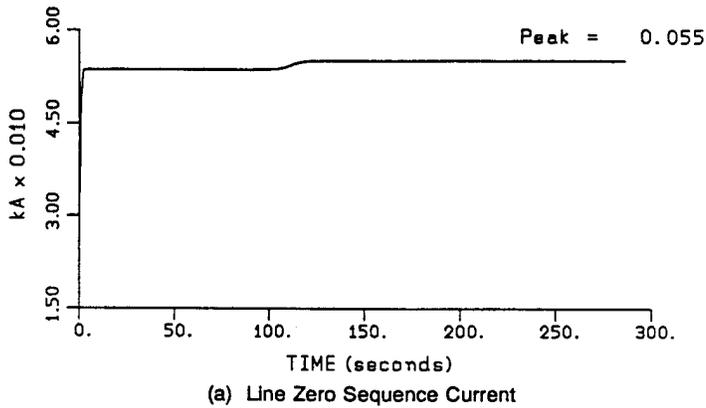

(a) Line Zero Sequence Current

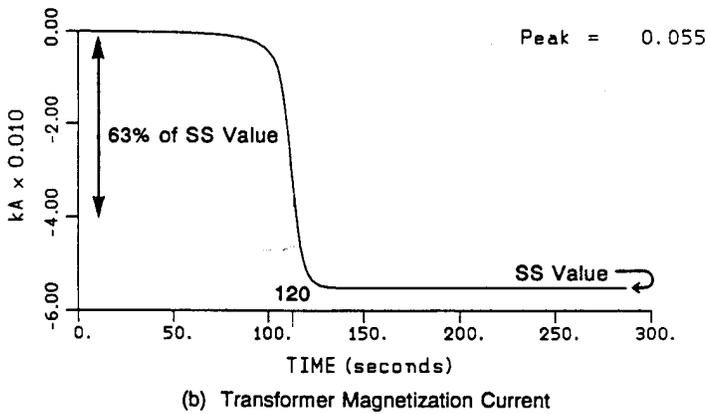

(b) Transformer Magnetization Current

GIV = 1 V/mi, Rtf = 5 Ohms, Rg = 4.435 Ohms/mi

Fig. 13. Time domain simulation results of the typical test system for : geomagnetically induced voltage = 1 V/mile; tower footing resistance = 5 ohms; and ground wire resistance 4.435 ohms/mile.

Transformer tests which have been performed at Minnesota Power have shown that the electric load has a substantial effect on transformer saturation time constants. This experimental result suggests that a proper power system model should exhibit the same behavior. To verify this fact, the model described in this paper has been utilized to study the effects of electric load on saturation time constants. For this purpose, a system similar to that of Fig. 10 with the following parameters was selected. GIV = 20 volts/km, soil resistivity 100 ohm-m, line length = 30 km, span length = 0. 1 miles, tower footing resistance = 25 ohms.

The GIV was assumed to be a step voltage. The transformer was assumed to be loaded with 0, 20%, 40%, and 100% three phase load which is wye connected. The same values of saturation time constants were obtained when the electric load was assumed to be delta connected. The computed saturation time constants are:

Transformer Load                                    Saturation Time Constant

| | 0% | 48 seconds |
| | 20% | 17 seconds |
| | 40% | 11 seconds |
| | 100% | 6 seconds |

|  | GIV | | | | | |
|---|---|---|---|---|---|---|
|  | 1 V/mile | | 10 V/mile | | 100 V/mile | |
| $R_t$ \ $R_g$ | 4.435 | 1.24 | 4.435 | 1.24 | 4.435 | 1.24 |
| 5 | 120 | 145 | 13.0 | 16.0 | 2.8 | 3.1 |
| 30 | 98 | 109 | 11.2 | 12.5 | 2.5 | 2.6 |
| 100 | 93 | 96 | 10.6 | 11.4 | 2.4 | 2.5 |
| ∞ | 85 | 85 | 10.0 | 10.0 | 2.3 | 2.3 |

*Time Constant to Saturation* is defined as the time required for the transformer magnetization current to reach 63.2% of its steady state value.

Table 3. Test system time constant to saturation* (in seconds) versus GIV level, tower footing resistance ($R_t$), and ground wire resistance ($R_g$).

The observed results can be explained with a simplified model of the zero sequence network. Specifically one can think of the saturation time constant as determined by an equivalent *R-L* circuit in the model. In the equivalent model, the inductance is dominated by the transformer inductances and the resistance is dominated by the transformer winding resistance and partly from the load resistance. The transformer magnetizing inductance can be thought of as being connected in parallel. The influence of the electric load on the equivalent resistance of the zero sequence model can be computed with standard circuit analysis methods. As an example, assuming a 20% electric load for a transformer with leakage impedance of 0.001 +j0.10 pu, the effect of the electric load is to double the equivalent resistance from 0.001 to 0.002 pu. Doubling the equivalent resistance will cause a substantial decrease of the saturation time constant due to the nonlinear characteristic of the magnetization inductance.

## 7. Comparison of the Effects of MHD-EMP and Solar Storms

The relative effects of MHD EMP and solar storm geomagnetically induced currents *(SS-GIC)* on power systems can be ascertained by the level of saturation reached due to typical values of geomagnetically induced voltages *(GIV)* from MHD EMP or solar storms. The maximum levels of saturation will be computed for the parameter values listed in Table 2. For this test the following two comparable strength MHD EMP and *SS-GIV* will be assumed:

Case 1. The geomagnetically induced voltage due to MHD EMP has a time variation as in Fig. 6 and a peak value of 100 volt/mile. The geomagnetically induced voltage due to a solar storm is practically dc and has a maximum value of 10 volt/mile.

Case 2. Same as in Case I except that the peak values are 10 and 1 volt/mile, respectively.

The results are illustrated in Tables 4 and 5. Note that for Case 2, even with the MHD EMP *GIV* ten times higher than the SS *GIV,* the maximum saturation level is comparable. For Case 1, the saturation level is much higher for MHD EMP excitation. In order to evaluate the significance of these results, it is important to recognize that the dc current through the transformer causes increased harmonic currents. Transformer losses increase causing the operating temperature of the transformer to rise. To determine the temperature rise, one must couple the electric model of the transformer with its thermal model. Such an integrated model has been developed by Meliopoulos and Cokkinides and validated in the laboratory. [28,29] Figure 14 illustrates the temperature response of a distribution transformer to step changes in the electric current through the transformer. Note that the time constants are quite long. We have determined that the increased losses in the transformer windings alone increase about 20% of full load losses for each 10% of dc current through the transformer, using the transformer model and typical values.

Table 4. Maximum dc offset magnetic flux (in pu) in transformer core: Case 1.

| $R_t \quad R_g$ | SS-GIV = 1 V/mile | | MHD-EMP-GIV = 10 V/mile | |
|---|---|---|---|---|
| | 4.435 | 1.25 | 4.435 | 1.25 |
| 5 | 0.129 | 0.128 | 0.142 | 0.136 |
| 30 | 0.130 | 0.129 | 0.143 | 0.140 |
| 100 | 0.130 | 0.130 | 0.144 | 0.143 |
| ∞ | 0.130 | 0.130 | 0.146 | 0.146 |

Table 5. Maximum dc offset magnetic flux (in pu) in transformer core: Case 2.

| $R_t \quad R_g$ | SS-GIV = 10 V/mile | | MHD-EMP-GIV = 100 V/mile | |
|---|---|---|---|---|
| | 4.435 | 1.25 | 4.435 | 1.25 |
| 5 | 0.147 | 0.143 | 0.292 | 0.270 |
| 30 | 0.150 | 0.147 | 0.300 | 0.281 |
| 100 | 0.151 | 0.149 | 0.310 | 0.298 |
| ∞ | 0.153 | 0.153 | 0.325 | 0.325 |

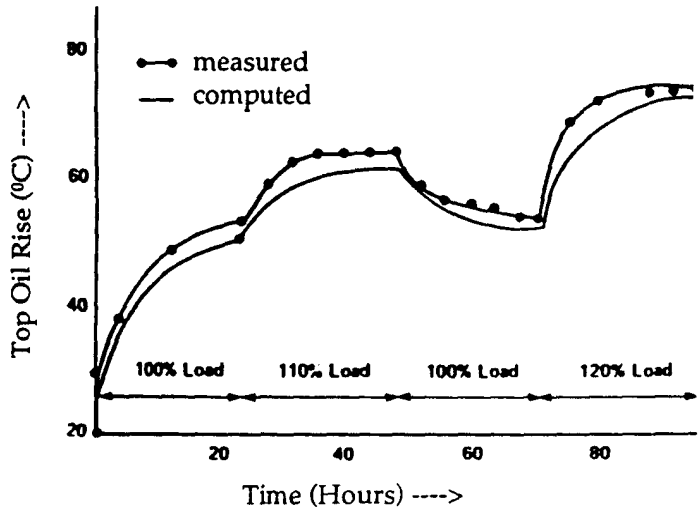

Fig. 14. Temperature response of transformer oil, cf. Ref. 29.

## 8. Conclusions

Since the power system periodically copes fairly well with solar storms which have an eleven year cycle related to sunspot activity, an important question we considered is how well it would cope with MHD EMP from a nuclear burst. For long transmission lines, i.e. 300 miles or longer, the dc-like induced current from geomagnetic perturbations offsets the 60 Hz ac and may saturate transformer cores, with secondary results such as high magnetization currents, increased harmonics, and concomitant effect on power system operation. The level of the transformer core saturation depends on the time constant of the saturation process, and on the duration and magnitude of the direct current through the transformer windings. Thus, models of transmission lines which explicitly represent grounding, earth potential, and frequency dependent phenomena, and power transformers with explicit representation of nonlinear magnetization characteristics were utilized.

Although the occurrence of substorms is by far the most frequent of the three geomagnetic perturbation phenomena, and their size dwarfs MHD EMP, the effects of substornis are relatively negligible on the earth. Comparative studies of MHD EMP and SS-GIV indicate'that (1) for low values of GIV, an MHD EMP level of 10 times the SS-GIV will cause comparable effects, and (2) for high values of GIV, the short duration of the MHD EMP mitigates the results only marginally. Since transformer thermal time constants are much larger than the duration of MHD EMP, the general conclusion is that the thermal effects due to MHD EMP induced voltages on power systems are less severe than those due to solar storm GIV. Shorter lines will be affected less.

With techniques developed in this paper, a parametric analysis of saturation time constants was performed and the effects of MHD EMP and SS GIC compared. A comprehensive model for studying the effects of geomagnetically induced voltages on power systems has been presented. The model has been used to study the time constants involved in reaching transformer saturation due to GIV (geomagnetically induced voltages) and the saturation level versus system parameters for a 473-mile long line. The most important parameters determining the effects of GIV on power systems are:

1. Level of GIV
2. Duration of GIV
3. Line length
4. Tower grounding impedance
5. Ground wire resistance

Based upon our analysis, we reached the following additional conclusions:

*Heating:* Heating is probably not a major consideration for GIV excitation due to MHD EMP because of its relatively short duration of high dc offset flux. As a matter of fact, saturation in this case lasts for about 100 seconds. On the other hand, GIV excitation due to solar storms lasts several hours. In this case, heating becomes a major consideration.

*Equipment shutdown*: Due to GIV excitation (SS or MHD EMP), transformers become generators of harmonics and absorbers of reactive power. Protective relaying, sensing this situation, may trip the transformer. In this case, irrespective of the duration of the GIV excitation, reversible equipment trips will occur with possibly major consequences such as the Hydro-Quebec blackout.

**Acknowledgments**

The authors would like to acknowledge the help of Dr. V. Albertson and Mr. J. Kappemann for providing the data of the test system. The discussions, inputs, and suggestions provided by Messrs. R. Barnes, F. Tesche, and R. Walling are appreciated. The investigation of the effects of load on saturation time constants was suggested by R. Barnes. This work was supported by the Electric Power Research Institute.